\documentclass[a4paper]{article}
\usepackage{comment}
\usepackage{url}
\usepackage{color}

\usepackage{INTERSPEECH2021}

\title{Improving robustness of one-shot voice conversion with deep discriminative speaker encoder}
\name{Hongqiang Du, Lei Xie{*}\thanks{*Lei Xie is the corresponding author, lxie@nwpu.edu.cn.}}
\address{
  Audio, Speech and Language Processing Group (ASLP@NPU), School of Computer Science, Northwestern Polytechnical University, China
  }
\email{hqdu@nwpu-aslp.org, lxie@nwpu.edu.cn}

\begin{document}

\maketitle
\begin{abstract}
One-shot voice conversion has received significant attention since only one utterance from source speaker and target speaker respectively is required. Moreover, source speaker and target speaker do not need to be seen during training. However, available one-shot voice conversion approaches are not stable for unseen speakers as the speaker embedding extracted from one utterance of an unseen speaker is not reliable. In this paper, we propose a deep discriminative speaker encoder to extract speaker embedding from one utterance more effectively. Specifically, the speaker encoder first integrates residual network and squeeze-and-excitation network to extract discriminative speaker information in frame level by modeling frame-wise and channel-wise interdependence in features. Then attention mechanism is introduced to further emphasize speaker related information via assigning different weights to frame level speaker information. Finally a statistic pooling layer is used to aggregate weighted frame level speaker information to form utterance level speaker embedding. The experimental results demonstrate that our proposed speaker encoder can improve the robustness of one-shot voice conversion for unseen speakers and outperforms baseline systems in terms of speech quality and speaker similarity.

\end{abstract}

\section{Introduction}

Voice conversion (VC) is a technique to modify the speech signal of a source speaker to make it sound like that of a target speaker without changing the linguistic content~\cite{mohammadi2017overview}. This technique has many applications, including expressive speech synthesis, speech enhancement, movie dubbing as well as other entertainment applications.

Various approaches have been proposed to achieve voice conversion, such as Gaussian
mixture model (GMM)~\cite{benisty2011voice,stylianou1998continuous,toda2007voice}, frequency warping approaches~\cite{erro2009voice,godoy2011voice,tian2015sparse}, exemplar based methods~\cite{takashima2012exemplar,wu2014exemplar,tian2017exemplar}, and neural network based methods~\cite{sun2015voice,hsu2016voice,kaneko2017parallel,wang2021accent,tian2020nus}. While these works require to know either source speaker or target speaker or both in training, which limits their use in the real application scenarios. Recently, one-shot voice conversion approaches~\cite{liu2018voice,lu2019one,chou2019one,qian2019autovc} are proposed. Compared with previous methods, source and target speakers at run-time are not required to be seen during training. Additionally, only one utterance from the source speaker and target speaker respectively is needed. The speaker identity of converted speech can be controlled independently by the speaker embedding extracted from target speech. 


Despite recent progress, the available one-shot voice conversion approaches are not stable for unseen speakers~\cite{huang2020far}. This is mainly because speaker embedding extracted from one utterance of an unseen speaker is not reliable~\cite{li2018deep,huang2020far}, which has a great influence on the stability of one-shot conversion~\cite{huang2020far}. Speaker embedding extractor can be a speaker encoder which is jointly trained with a conversion model or a pre-trained model for speaker information extraction, such as i-vector~\cite{dehak2010front}, d-vector~\cite{doddipatla2017speaker}, or x-vector~\cite{snyder2018x}. The speaker embedding extractor jointly trained with the conversion model is more suitable for voice conversion than the pre-trained models~\cite{huang2020far}. When the network is jointly optimized, speaker embedding extractor is an inherent part of the model, which makes the generation of speech with correct speaker embedding consistently.





There are some studies on jointly training speaker encoder and voice conversion model. The speaker encoder generally consists of two parts: extracting frame level and utterance level features~\cite{okabe2018attentive}. The frame level extractor takes acoustic features as input and outputs frame level features. It can be done via recurrent neural networks~\cite{zhou2019many} and convolutional neural networks~\cite{chou2019one,lu2019one}. In particular, convolutional neural network based residual network (ResNet) is a powerful speaker embedding extractor~\cite{gusev2020deep,chou2019one}. The utterance level extractor further aggregates variable-length frame level features into utterance level speaker embedding. Average pooling~\cite{zhou2019many,chou2019one} is a popular method to obtain speaker embedding by averaging all frame level features. Another method~\cite{lu2019one} first uses the last state of unidirectional gated recurrent unit (GRU) layer as the utterance level speaker representation and then multi-head attention is utilized as a post-processing module to obtain final speaker embedding. 




In this paper, to further improve the effectiveness of speaker embedding extracted from only one utterance of an unseen speaker, we propose a deep discriminative speaker encoder. Inspired by~\cite{xu2020deep}, first residual network and squeeze-and-excitation network~\cite{hu2018squeeze} are integrated to extract discriminative frame level speaker information by modeling frame-wise and channel-wise interdependence in features. Then attention mechanism is introduced to give different weights to frame level speaker information. Finally, a statistic pooling layer~\cite{snyder2017deep} is used to aggregate weighted frame level speaker information to generate utterance level speaker embedding.  Experimental
results show that our proposed speaker encoder can improve the robustness of one-shot voice conversion and outperforms baseline systems in terms of  speech quality and speaker similarity.



\section{Related work}
\label{sec:related_work}

\subsection{Residual network based speaker embedding extractor}


Residual network (ResNet) has been widely used in speaker verification, which achieves promising performance for both long-duration and short-duration utterances~\cite{li2018deep,gusev2020deep}.

%

The architecture of ResNet based speaker embedding extractor includes several ResNet blocks, followed by a statistics pooling layer and fully-connected (FC) layers. ResNet block operates on frame level features, which consists of convolutional layers, rectified linear units (ReLU) and batch normalization (BN) layers. Residual connection in ResNet block helps to build a deep neural network and avoids the vanishing gradient problem~\cite{he2016deep,garcia2020jhu}. Increasing the depth of a neural network can significantly improve the quality of representations~\cite{hu2018squeeze}. Additionally, batch normalization helps to improve the stability of the training process of deep neural networks.  Then a statistics pooling layer calculates the mean and standard deviation of each sample along the time-axis to form utterance level representation. Finally, two fully-connected (FC) layers project the utterance level representation into a fixed dimensional speaker embedding.

\subsection{Squeeze-and-excitation network}

Squeeze-and-excitation (SE) network~\cite{hu2018squeeze} is first introduced to model channel interdependence in features for image classification. SE network can be used as a block and inserted in the convolutional neural network.

SE block consists of two operations: squeeze operation and excitation operation. Squeeze operation utilizes average pooling to generate channel-wise statistics. The statistics are mean vector $z$ of frame level features ${h_t}$ across the time-axis.

\begin{equation}\label{eq:squeeze}
	z = \frac{1}{T}\sum\limits_t^T {{h_t}}
\end{equation}

Then $z$ is used in the excitation operation to calculate weights for each channel. The excitation operation is formulated as follows:

\begin{equation}\label{eq:excitation}
s = \sigma ({W_2}\delta ({W_1}z))
\end{equation}
where $\sigma ( \cdot )$ and $\delta ( \cdot )$ are sigmoid and ReLU function respectively. ${W_1} \in {R^{C \times \frac{C}{r}}}$, ${W_2} \in {R^{\frac{C}{r} \times C}}$, $C$ and $r$ refer to the number of input
channel and reduction ratio respectively. The channel-wise vector $s$ contains the weight for each channel, which is between zero and one.

The final output ${\hat h}$ of SE block is obtained by channel-wise multiplication between the original input $h$ and corresponding weight in $s$. 

\begin{equation}\label{eq:final}
\hat h = sh\
\end{equation}

\section{Robust one-shot voice conversion}
\label{sec:proposed}


\subsection{Deep discriminative speaker encoder}

Speaker encoder is an important component in the framework of one-shot voice conversion, which is directly related to performance of the whole network for unseen speakers~\cite{huang2020far}. 

\begin{figure*}[!ht]
	\centering
	\centerline{\includegraphics[width=1.0\linewidth]{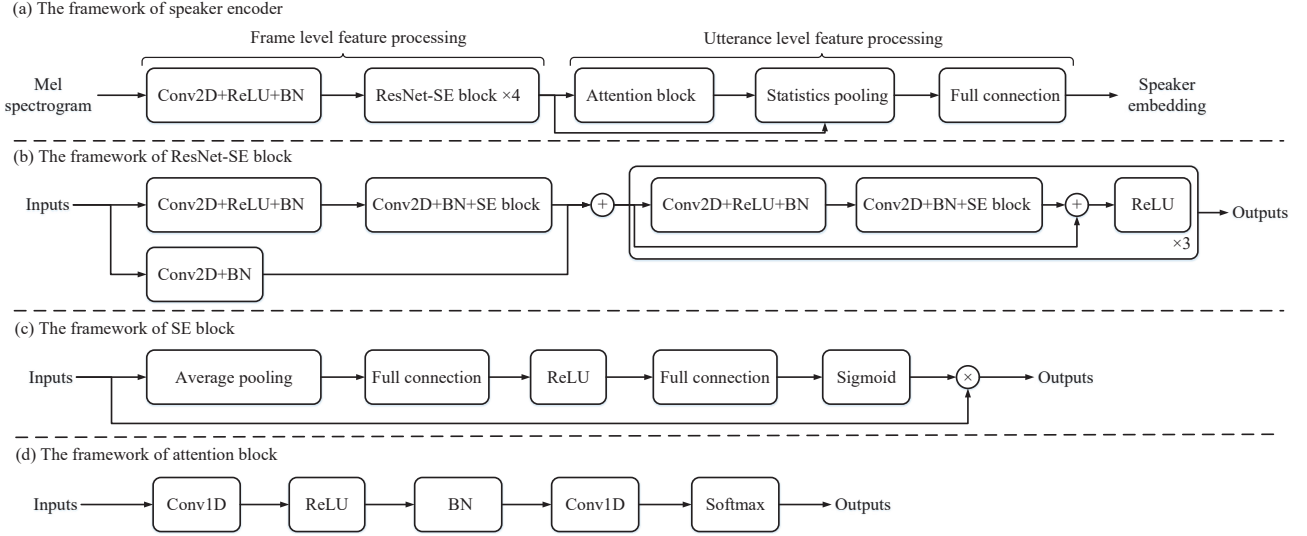}}
	\caption{The framework of the proposed deep discriminative speaker encoder for robust one-shot voice conversion. The speaker encoder consists of frame level feature processing and utterance level feature processing. }
	\label{fig:speaker-encoder}
\end{figure*}

Inspired by the previous study on ResNet and SE block, we integrate ResNet with SE block to build a deep discriminative speaker encoder (DDSE) for robust one-shot voice conversion. Figure~\ref{fig:speaker-encoder} (a) depicts the framework of speaker encoder. It consists of frame level feature processing and utterance level feature processing. 


The frame level feature processing part consists of a Conv2D layer, a ReLU function and a batch normalization (BN) layer, followed by four ResNet-SE blocks. The framework of ResNet-SE block is shown in Figure~\ref{fig:speaker-encoder} (b). A ResNet-SE block mainly consists of convolution layers. Filters in the convolution layer explicitly model local features 
and allow spatial translation invariance, which make convolution layer suitable to extract frame level features~\cite{xu2020deep}. SE block expands the temporal context of the frame level information by modeling channel interdependence in features, which has been verified to be helpful in speaker verification task~\cite{xu2020deep}. The framework of SE block is shown in Figure~\ref{fig:speaker-encoder} (c). An average
pooling layer is utilized to generate channel-wise statistics. Then, two
fully-connected layers capture the local channel dependencies. The first fully-connected  layer can be used to reduce the feature dimension for controlling the computational cost, while the second fully-connected  layer restores the number of feature to the original dimension. Finally, the channel-wise vector is obtained with a sigmoid layer to  pay more attention to the discriminative channels for speaker representation.


The utterance level feature processing part consists of an attention block, followed by a statistic pooling layer and two fully-connected layers. Instead of directly using an average pooling layer where each frame level speaker information contributes equally to speaker embedding~\cite{chou2019one}, we first introduce an attention block to further emphasize speaker related information. As shown in Figure~\ref{fig:speaker-encoder} (d), the attention block takes the  frame level speaker information as input and outputs the corresponding  weights, which allows the speaker encoder to select the frames it deems relevant. Then a statistics pooling layer~\cite{okabe2018attentive} is used to calculate the weighted means and weighted standard deviations of the final extracted frame level
features. The mean and deviation are combined to form an utterance-level speaker representation. Finally, two
fully-connected layers are introduced. The first one acts
as a bottleneck layer to generate the low-dimensional speaker
representation. The second one projects the speaker
representation into a fixed dimensional speaker embedding.

In summary, convolution based ResNet is a powerful architecture to extract speaker representations by modeling relationships between frames. SE block contributes to the
discriminative speaker representation learning by exploring the channel-wise information. Attention mechanism further makes speaker encoder emphasize speaker related information and overshadow other information. Therefore, the speaker
embedding learned by this architecture concentrates on speaker characteristics more effectively.


\subsection{Robust one-shot voice conversion with DDSE}

%


The frameworks of available one-shot voice conversion approaches~\cite{chou2019one,qian2019autovc} generally consist of a speaker encoder, a content encoder, and a decoder.  AdaIN-VC~\cite{chou2019one} is a successful end-to-end implementation, which is relatively more robust for unseen speakers~\cite{huang2020far}. We use AdaIN-VC as a case study and replace the original speaker encoder with our proposed deep discriminative speaker encoder to make it more robust for unseen speakers. Note that the whole network is jointly optimized and the speaker encoder is optimized without explicit loss function.

\begin{figure}[!ht]
	\centering
	\centerline{\includegraphics[width=1.0\linewidth]{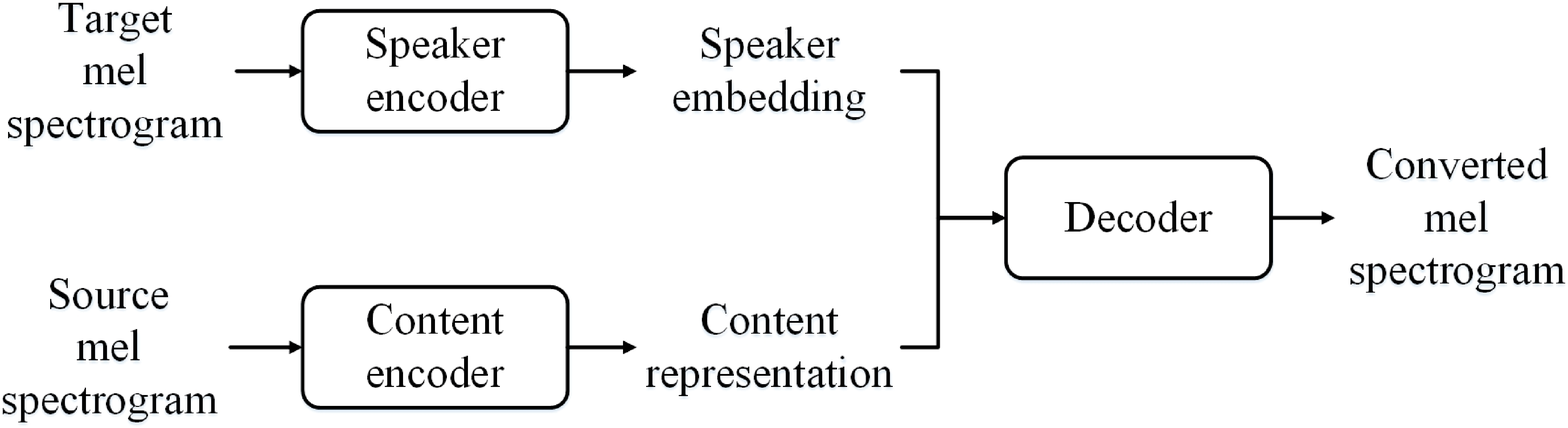}}
	\caption{The diagram of robust one-shot voice conversion at run-time conversion.}
	\label{fig:general}
\end{figure}

The robust one-shot voice conversion method consists of two steps: conversion model training and run-time conversion. During the training stage, the speaker encoder and content encoder learn to extract speaker embedding and linguistic content representation from spectrum respectively. The decoder takes the two representations as inputs to reconstruct the spectrum. During run-time shown in Figure~\ref{fig:general}, the speaker encoder extracts utterance level speaker embedding from target speech. The content encoder extracts frame level content representation from source speech. The decoder takes content and speaker representations as input to reconstruct converted speech.

\section{Experimental setup}
\label{sec:experiment setups}
\subsection{Database and feature extraction}
\label{ssec:Database}

CSTR-VCTK~\cite{veaux2017cstr} database, containing 44 hours of speech from 109 speakers, is used to train the conversion model. Voice conversion experiments are carried out on CMU-ARCTIC~\cite{kominek2004cmu} database and VCC 2016 dataset~\cite{toda2016voice} respectively. For CMU-ARCTIC database, we select clb (female) and rms (male) as source speakers, and slt (female) and bdl (male) as target speakers. For VCC 2016 dataset, SF1 (female) and 
SM1 (male) are selected as source speakers, and TF2 (female)
and TM3 (male) are selected as target speakers. For each target speaker, 20 utterances are used for evaluation. All audio files are downsampled to 16 kHz.


Librosa is employed to extract 256 dimensional mel spectrogram with 50ms frame length and 12.5ms frame shift.

\subsection{Systems and setup}
\label{ssec:Baselines_setup}

\begin{itemize}
\item VAE-ORI: This is the original AdaIN-VC~\cite{chou2019one} system. The speaker encoder and content encoder take 256 dimensional mel spectrogram as input and the output is 128 dimensional speaker embedding and content representation respectively. To further improve the speech quality, the auto-regressive technique is applied to the decoder. 
\item VAE-GSE: This has the same setting as VAE-ORI except that the speaker encoder is replaced with global speaker embedding (GSE) utilized in~\cite{lu2019one}. 
\item VAE-ResNet: This has the same setting as VAE-ORI except that the speaker encoder is replaced with ResNet. 
\item VAE-DDSE: This has the same setting as VAE-ORI except that the speaker encoder is replaced with our proposed deep discriminative speaker encoder (DDSE).  For the first ResNet-SE block, the kernel sizes and strides for the Conv2D layers are ${\rm{\{ 3, 3, 1, 3, 3\} }}$ and ${\rm{\{ \{ 1,1\} , \{ 1,1\} , \{1,1\}, \{ 1,1\}, \{ 1,1\} \} }}$ respectively. For the remaining ResNet-SE blocks, the kernel sizes and strides for the Conv2D layers are ${\rm{\{ 3, 3, 1, 3, 3\} }}$ and ${\rm{\{ \{ 2,2\} , \{ 1,1\} , \{2,2\}, \{ 1,1\}, \{ 1,1\} \} }}$ respectively. For the SE block, the reduction ratio $r$ is set to 8. 
\end{itemize}


Parallel WaveGAN~\cite{yamamoto2020parallel} is used to synthesize the converted speech. We follow the original configurations.

\begin{figure*}[!ht]
	\centering
	\centerline{\includegraphics[width=1.0\linewidth]{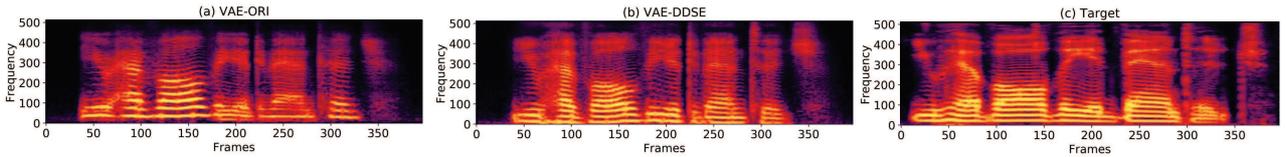}}
	\caption{Example of spectrum of the same utterance converted from clb (female) to bdl (male) by different systems: (a) VAE-ORI, (b) VAE-DDSE, and (c) Target.}
	\label{fig:spectrum}
\end{figure*}

\section{Evaluations}
\label{sec:experiment results}
%
%

\subsection{Objective evaluation}
\label{ssec:subhead}

Mel-cepstral distortion (MCD)~\cite{toda2007voice} is employed to measure the spectral distortion. MCD is the Euclidean distance of the mel spectrogram between the converted speech and the
target speech. Given a speech frame, the MCD is defined as follows: 
\begin{equation}\label{eq:mcd}
	\vspace{1mm}
	\text{MCD[dB]} = \frac{{10}}{{\ln 10}}\sqrt {2\sum\limits_{n=1}^N {{{\left( {X_n^{conv} - X_n^{ref}} \right)}^2}} },
\end{equation}
where $X_n^{conv}$ and $X_n^{ref}$ are the $n^{th}$ coefficient of the converted and target mel spectrogram, $N$ is the dimension of mel spectrogram. The lower MCD indicates the smaller distortion.

Table~\ref{table:mcd} shows the average MCD for different systems.  Intra-gender conversion is better than inter-gender conversion for the four systems. We also observe that VAE-DDSE significantly outperforms the VAE-ORI and VAE-GSE in both intra- and inter-gender conversions. VAE-DDSE performs better than VAE-ResNet and achieves the lowest average MCD of 10.75 dB. The objective evaluation results further confirm that the extracted speaker embedding has a great impact on the performance of one-shot voice conversion.

\begin{table}[!h]
	\caption{Comparison of average MCD (dB) for different systems.}
	\small
	\renewcommand\arraystretch{1.5}
	\centering
	\setlength{\tabcolsep}{4.5mm}{
		\begin{tabular}{cccc}
			\hline
			System   & Inter & Intra & Average \\ \hline
			VAE-ORI  & 13.02 & 12.85 & 12.93   \\ 
			VAE-GSE  & 15.66 & 15.52 & 15.59   \\ 
			VAE-ResNet & 10.95 & 10.81 & 10.88 \\
			VAE-DDSE & 10.81 & 10.70 & 10.75   \\ \hline
	\end{tabular}}
	\label{table:mcd}
\end{table}

In Figure~\ref{fig:spectrum}, we show an example that compares spectrum  of the same utterance converted from clb (female) to bdl (male) by different systems: (a) VAE-ORI, (b) VAE-DDSE, and (c) Target. The harmonics of the spectrum are closely related to the speaker identity~\cite{hsu2017learning}, which is controlled by the extracted speaker embedding. The harmonics in Figure~\ref{fig:spectrum} (a) are clearly higher than that in Figure~\ref{fig:spectrum} (c), which indicates that the converted speech is not stable and speaker similarity is degraded. The harmonics maintain roughly the same between Figure~\ref{fig:spectrum} (b) and Figure~\ref{fig:spectrum} (c).



\subsection{Subjective evaluation}
\label{sssec:subsubhead}

For subjective evaluation, we first conduct AB and XAB preference tests to assess speech quality and speaker similarity. Then the mean opinion score (MOS) is utilized to evaluate speech naturalness. Each listener is asked to give an opinion score on a five-point scale (5: excellent, 4: good, 3: fair, 2: poor, 1: bad).
For each system, 20 samples are randomly selected from the 160 converted samples for listening tests. Ten listeners participated in all listening tests. Different listeners may listen to different samples. Listening tests cover all the 160 evaluation samples.

\begin{figure}[!ht]
	\centering
	\centerline{\includegraphics[width=1.0\linewidth]{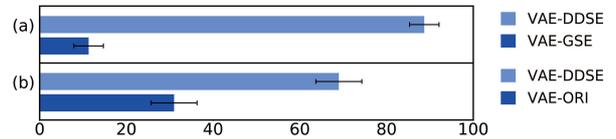}}
	\caption{Quality preference tests of converted speech samples with 95\%	confidence intervals for different systems.}
	\label{fig:ab}
\end{figure}

The subjective results of AB tests are presented in Figure~\ref{fig:ab}. It is observed that our proposed VAE-DDSE significantly outperforms VAE-GSE and VAE-ORI. 

\begin{figure}[!ht]
	\centering
	\centerline{\includegraphics[width=1.0\linewidth]{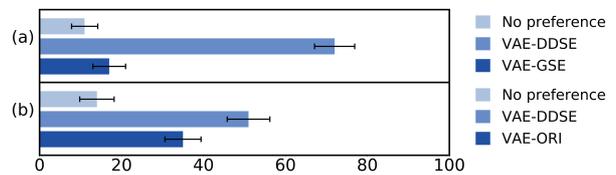}}
	\caption{Similarity preference tests of converted speech samples with 95\% confidence intervals for different systems.}
	\label{fig:abx}
\end{figure}

Figure~\ref{fig:abx} shows the similarity preference results of XAB
tests. As shown in Figure~\ref{fig:abx} (a) and (b), we observe that VAE-DDSE outperforms VAE-GSE and VAE-ORI respectively in terms of speaker similarity.

\begin{figure}[!ht]
	\centering
	\centerline{\includegraphics[width=1.0\linewidth]{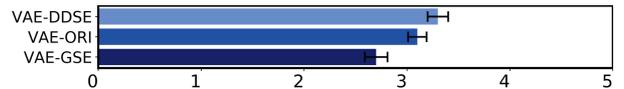}}
	\caption{Comparison of mean opinion scores with 95\% confidence intervals for different systems}
	\label{fig:mos}
\end{figure}

Figure~\ref{fig:mos} shows the mean opinion scores for different systems. Benefiting from the discriminative speaker embedding extracted from DDSE, VAE-DDSE achieves the highest MOS score. The synthesized samples can be found on the website~\footnote{\url{https://dhqadg.github.io/robust/}}.

\section{Conclusions}
\label{sec:conclusion}
In this study, we propose a deep discriminative speaker encoder to improve the robustness of one-shot voice conversion for unseen speakers. The speaker encoder first integrates residual network and  squeeze-and-excitation network to extract frame level speaker information from time-axis and channel-axis. Then attention mechanism is used to further focus on the speaker related information. Finally a statistic pooling layer is used to aggregate weighted frame level speaker information to form utterance level speaker embedding. The experimental results demonstrate that our proposed speaker encoder can improve the robustness of one-shot voice conversion for unseen speakers in terms of speech quality and speaker similarity.


\bibliographystyle{IEEEtran}

\bibliography{interspeech}


\end{document}